\documentclass[traditabstract]{aa}
\usepackage{graphicx}
\usepackage{txfonts}
\usepackage{epsfig}
\usepackage{natbib}
\usepackage{color}
\usepackage{rotating}
\usepackage[draft,pagebackref=true]{hyperref}
\hypersetup{colorlinks=true,linkcolor=blue,citecolor=blue,filecolor=magenta,urlcolor=blue}
\usepackage{ulem}
\usepackage[dvipsnames]{xcolor}

\newcommand{\oversim}[2]{\protect{\mbox{\lower0.5ex\vbox{%
  \baselineskip=0pt\lineskip=0.2ex
  \ialign{$\mathsurround=0pt #1\hfil##\hfil$\crcr#2\crcr\sim\crcr}}}}}
\usepackage{lineno}
\begin{document}

\title{Evidence for feedback and stellar-dynamically regulated bursty
  star cluster formation: the case of the Orion Nebula Cluster}
\author{Pavel Kroupa\inst{1,2}, Tereza Je\v{r}\'{a}bkov\'{a}\inst{1,2,3},
  Franti\v{s}ek Dinnbier\inst{4}, Giacomo Beccari\inst{3}, Zhiqiang Yan\inst{1,2}
}
\institute{
Helmholtz-Institut f\"ur Strahlen- und Kernphysik, University
        of Bonn, Nussallee 14-16, D-53115 Bonn, Germany\\
\email{pavel@astro.uni-bonn.de}
\and Charles University in Prague, Faculty of Mathematics and 
        Physics, Astronomical Institute, V Hole\v{s}ovi\v{c}k\'ach 2, CZ-180
        00 Praha 8, Czech Republic\
\and European Southern Observatory, Karl-Schwarzschild-Str. 2, 85748
Garching, Gernamy\\ \email{tjerabko@eso.org}
\and 
1 Physikalisches Institut, Universit\"at zu K\"oln, Z\"ulpicher Str. 77, 50937 K\"oln, Germany\\
\email{dinnbier@ph1.uni-koeln.de}
}

\date{accepted 28.12.2017} 

\abstract{A scenario for the formation of multiple co-eval populations
  separated in age by about 1~Myr in very young clusters (VYCs, ages
  less than 10~Myr) and with masses in the range $600-20000\,M_\odot$
  is outlined. It rests upon a converging inflow of molecular gas
  building up a first population of pre-main sequence stars. The
  associated just-formed O stars ionise the inflow and suppress star
  formation in the embedded cluster.  However, they typically eject
  each other out of the embedded cluster within $10^6\,$yr, that is
  before the molecular cloud filament can be ionised entirely. The
  inflow of molecular gas can then resume forming a second
  population. This sequence of events can be repeated maximally over
  the life-time of the molecular cloud (about 10~Myr), but is not
  likely to be possible in VYCs with mass $<300\,M_\odot$, because
  such populations are not likely to contain an O~star.  Stellar
  populations heavier than about $2000\,M_\odot$ are likely to have
  too many O stars for all of these to eject each other from the
  embedded cluster before they disperse their natal cloud.  VYCs with
  masses in the range $600-2000\,M_\odot$ are likely to have such
  multi-age populations, while VYCs with masses in the range
  $2000-20000\,M_\odot$ can also be composed solely of co-eval,
  mono-age populations.  More massive VYCs are not likely to host
  sub-populations with age differences of about 1~Myr.  This model is
  applied to the Orion Nebula Cluster (ONC), in which three
  well-separated pre-main sequences in the color-magnitude diagram of
  the cluster have recently been discovered.  The mass-inflow history is
  constrained using this model and the number of OB stars ejected from
  each population are estimated for verification using Gaia data.  As
  a further consequence of the proposed model, the three runaway O
  star systems, AE~Aur, $\mu$~Col and $\iota$~Ori, are considered as
  significant observational evidence for stellar-dynamical ejections
  of massive stars from the oldest population in the ONC.  Evidence
  for stellar-dynamical ejections of massive stars in the currently
  forming population is also discussed.

}

\titlerunning{Multiple populations in the ONC} 

\authorrunning{Kroupa et al.}  

\maketitle

\section{Introduction}
\label{sec:introd}

The Orion Nebula Cluster (ONC) is about 400~pc distant from the Sun
\citep{Menten+07, Kounkel+17}, approximately $1-3\,$Myr old, has a
current stellar mass of $\approx 1500\,M_\odot$ partially embedded in
molecular gas within a current half-mass radius of approximately 0.8~pc and a
nominal cluster radius near $2.5\,$pc \citep{HH98, DaRio+09,
  Megeath16, DaRio+17}
and has been understood to be the nearest massive young stellar
cluster allowing much detailed observational work, including the gas
flows within and out of it \citep{ODell+09, McLeod+16,Hacar+17}. The
inner ONC has been known as the Trapezium cluster \citep{Prosser+94}.
Within the nominal radius of the ONC, intense star-formation is still
on-going \citep{ODell+09} and the violent stellar-dynamical ejection
of massive stars from this region has been mapped \citep{Tan04, ChT12,
  Luhman+17, FT18}.  The most massive O7V star, $\theta^1$C~Ori, is
currently ionising part of the molecular gas within the ONC cluster
and the Trapezium has been suggested to be astro-physically and
stellar-dynamically very young ($<10^5\,$yr old, \citealt{ODell98}).
$\theta^1$C~Ori is moving rapidly out of the Trapezium
\citep{ODell+09} suggesting that a stellar-dynamical encounter
happened recently \citep{Tan04, ChT12, FT18}.  The central Trapezium
of few massive stars has dimensions of only $0.05\,$pc and is
stellar-dynamically highly unstable with a short ($<10^5\,$yr)
life-time \citep{PAK06}.  The ONC has a top-light stellar initial mass
function (IMF) which would be consistent with a canonical IMF
\citep{Kroupa01, Kroupa13} if the cluster has efficiently expelled a
large fraction of its more-massive-than-$5\,M_\odot$ stars. That the
ONC is actively expelling stars stellar-dynamically is evident from
proper motion and radial velocity surveys \citep{Poveda+05}.
Stellar-dynamical modelling of the ONC suggests it may be expanding
due to the expulsion of residual gas \citep{Kroupa+99, K2000, KAH,
  Scally+05}.  This is consistent with the overall evolution of
embedded clusters into the Galactic field population \citep{Lada10}
and the need for embedded clusters to rapidly (within a few~Myr)
expand to the radii of open clusters \citep{BK17}.

Observed very young and compact (half-mass radii smaller than 1~pc)
star clusters with ages between about 1 and 3~Myr and masses in the
range $10^3-10^5\,M_\odot$, such as the ONC, NGC3603 and R136, have
been understood to have formed essentially monolithically over a time
of about one~Myr \citep{KAH,Scally+05,BK13,BK14}.  This leaves little
time for them to be assembled through hierarchical merging of
pre-formed smaller clusters, because also the smooth spatial and
kinematical distributions of the observed stars need to be reproduced
\citep{Scally+02,BK15, BK15b}. 

The OmegaCAM survey by \cite{Beccari+17} finds the colour-magnitude
diagram of ONC stars to contain three well separated sequences.  An
interpretation offered by \cite{Beccari+17} based on these data is
that the ONC contains three dominant populations of stars.  Here we
outline how a star cluster might form in-situ in a series of formation
epochs (or `bursts', each burst forming a co-eval stellar population
monolithically, that is as one compact embedded cluster) from
molecular gas supplied by one and the same molecular cloud filament.
Recent modelling of the formation of star formation in
self-gravitating, initially turbulent and magnetised molecular clouds
shows that local star cluster formation in those simulations occurs on
time scales of the order of $10^5\,$yr associated with filamentary
accretion and locally converging gas flows (\citealt{Federrath15,
  Federrath16}, see also \citealt{Smith+16, Burkert17}). Assuming the
converging gas inflow along the molecular filament forms an embedded
cluster over a few$~10^5\,$yr (observational evidence: for example
\citealt{Schneider+12,Hacar+13, Hacar+17b}), star formation within the
embedded cluster can be stopped due to the ionisation of the inflowing
molecular filaments and blow-out of residual gas by a first generation
of~O stars. These are expelled from the cluster core through three-
and four-body encounters, allowing the inflow to resume and to build
up a next generation, provided the O stars did not disrupt the entire
molecular cloud before they ejected each other.

This model is thus based on a combination of termination of star
formation through O stars, the blow-out of the residual gas and
subsequent expansion of the already monolithically formed stellar
population \citep{KAH} and stellar-dynamical ejection of the ionising
stars from the cluster \citep{PAK06, OKP15, OK16} such that the cloud
is not disrupted and the cluster can resume accretion from the cold
molecular cloud \citep{PAK09}.  These authors explicitly calculated
(their sec.~6) that VYCs weighing down to $1000\,M_\odot$ can accrete
cold gas.  This process can repeat itself, such that this model
suggests a possible explanation of the time- and spatial-structure of
the stellar population(s) in the ONC as observed by
\cite{Beccari+17}. These observations may thus provide exquisite
evidence of how a stellar cluster is build-up by consecutive
gas-accretion phases subject to a fine level of self-regulation
through feedback which is interrupted by stellar-dynamical
processes. We emphasise that this may not be the only viable scenario
for explaining the apparent multiple age populations in the ONC. Here
we merely discuss and propose it as one of the possibilities.

In Section~\ref{sec:model} the model is described.
Section~\ref{sec:disc} provides a discussion with a few predictions,
and Sec.~\ref{sec:concs} ends with the conclusions.

\section{Model}
\label{sec:model}

\subsection{Preliminaries}

According to \cite{Beccari+17} the colour-magnitude diagram of the ONC
contains three well separated sequences, posing a challenge to the
theory of how such clusters form.  The authors note that the
interpretation that these may be due to unresolved multiple stellar
systems would require an unusual mass-ratio distribution not favoured
by the observed properties of multiple systems
\citep{Belloni+17}. This rejection needs to be treated with the caveat
that it may be possible that the youngest population evident in the
colour--magnitude diagramme may be the binary-star sequence of the
second oldest population, which by chance overlays the binary sequence
of the oldest population. The youngest population, rather than being
the youngest, could also be made up of triple stars of the oldest
population.  Since younger populations are more concentrated this
hypothesis would require the binary stars to have mass-segregated. The
stars mapped by \cite{Beccari+17} have masses in the range
$0.25-0.4\,M_\odot$.  The mass-segregation time-scale for
$0.7\,M_\odot$ systems is roughly~$4\,$Myr (Eq.~6 and~7 in
\citealt{Kroupa08}), making this an unlikely possibility, subject to
the additional constraint that the binary population ought to have a
realistic mass-ratio distribution
\citep{Beccari+17}. Stellar-dynamical modelling would be needed to
study this possibility, namely that the ONC is the outcome of a single burst
that happened 3~Myr ago, that is that it consists of only one population
of single, binary- and triple systems, instead of two and possibly
three populations separated by age.

The alternative interpretation that the younger populations may
constitute fore-ground is also excluded by \cite{Beccari+17} on the
grounds that the putative foreground population would need to be only
at a distance of about 200~pc from the Sun and by the spatial extend
of the stars correlating with their age.  Essentially, the apparently
younger, putative fore-ground population would show a comparable
spatial extend but it is observed that the concentration decreases
with age.  

That the older sequences may constitute captured populations from
pre-existing pre-main sequence stars (PMSs) in the region \citep{PAK07} may
be viable, as they would also be more spatially extended. However, in
this case the intermediate population would have to have had a smaller
velocity dispersion as it is also intermediately concentrated around
the ONC between the youngest and oldest.

Another interpretation offered by \cite{Beccari+17} is that the ONC
contains three dominant populations of stars with ages of
$10^{6.46\pm0.06}, 10^{6.27\pm0.09}, 10^{6.09\pm0.07}\,$yr,
corresponding to 2.88, 1.86 and 1.23~Myr.  It is important to note
that these ages were calculated in \cite{Beccari+17} as the mean ages
of the stars in the three discovered populations, were the ages of
each single star was taken form the age estimation given in
\cite{DaRio+16}.  An independent check using PMS isochrones-fitting
from \cite{Bressan+12} seems to show consistent median ages for the
oldest population but slightly younger median ages for the reddest
populations. The three populations would then show an age of 3, 1.8
and 0.8~Myr, respectively (Beccari, private communication).  

For the present contribution we assume the three PMS populations
discovered by \cite{Beccari+17} to be three age sequences
with these ages.  The youngest stars comprise the smallest group in
\cite{Beccari+17}, are the fastest rotators and are also more
centrally concentrated within the ONC than the older stars. Given the
young age it is postulated here that this youngest population is still
forming, because the ONC has, within its embedded molecular cloud
region, active on-going star-formation.  The oldest population
comprises the most massive group and is distributed around the ONC and
also follows the surrounding integral-shaped filament which appears to
feed the ONC with gas \citep{Hacar+17}.

\subsection{Ansatz}

Assuming that a molecular cloud forms a filament with a variable
density along its length, it follows that molecular gas flows from
both sides along the filament into the densest region which
constitutes the local potential well \citep{Burkert17}.
\cite{Hacar+17} have mapped the accelerated motion of molecular gas
into the ONC finding that the current inflow rate is about
$\dot{M_{\rm g}}=55\,M_\odot\,{\rm Myr}^{-1}$ per molecular finger
implying about $\dot{M_{\rm g}}=385\,M_\odot\,{\rm Myr}^{-1}$ for the
seven molecular fingers detected in OMC-1. Over $3\,$Myr the inflow,
if constant over this time, would accumulate a mass of
$1000\,M_\odot$. The inflow may have been larger in the past (for
example, the ionising star $\theta^1$C~Ori is currently affecting the
observed inflow of molecular material).  Also, the star formation
efficiency (SFE) on an embedded-cluster (one-pc) scale is
$\epsilon\approx 0.33$. This is found observationally \citep{LL03,
  Andre+14, Megeath16}, and the stellar-dynamical modelling of the
ONC, NGC3603 and R136 yield corroborative and consistent values also
\citep{BK15b}. Thus, a mass comparable in order of magnitude to the
observed mass of the ONC will have build-up.

The inflow builds up a population of PMSs, this taking typically a
few~$10^5\,$yr to a Myr. This can be deduced observationally in
forming embedded clusters (for example the Taurus-Auriga groups,
$\rho$~Oph, NGC~1333, \citealt{Hacar+13, Hacar+17b}).
\cite{Duarte-Cabral+13} find the formation times of high- and low-mass
stars cannot be distinguished, being about $10^5\,$yr. Simulations
also show that 95~per cent of the mass of a star is acquired in about
0.1~Myr (\citealt{WT03}, see also \citealt{Kuruwita+17}).  The fibers
of molecular cloud filaments observed by \cite{Hacar+17b}, in which
the newest generation of proto-stars is forming, can have a life-time
of only half a crossing time through~NGC1333 which is about~0.5-1~Myr.
The most-massive stars form in the densest regions near the centre of
the proto-cluster. This is observed to be the case \citep{Bontemps+10,
  Lane+16, Kirk16}, and star-formation simulations also show
primordial mass segregation \citealt{MC11, BS11}).  Also, it is
observed that the more massive the embedded cluster is, the more
massive is the most massive PMS in it \citep{Megeath16, Ramirez16,
  Stephens+17, Yan+17}. Monolithically-formed embedded clusters
typically have radii smaller than a~pc \citep{MK12}.

The radiation of stars more massive than about $8\,M_\odot$ ionises
the inflow. Further inflow of molecular star-forming gas into the
inner region of the cluster may thus be shut-off once the embedded
cluster spawns ionising stars. Star-formation may continue within the
molecular filament such that not all star formation is necessarily
stopped, but the star-formation rate in the embedded cluster is likely
to be reduced in the presence of ionising stars. Stars formed in the
filaments may orbit through the embedded cluster but will reach to
larger radii on the other side because they decouple from the
hydrodynamics and become ballistic particles on a time-scale of about
$10^5\,$yr \citep{WT03,Duarte-Cabral+13}.

\begin{figure*}
	\begin{center}
		\includegraphics[scale=0.75]{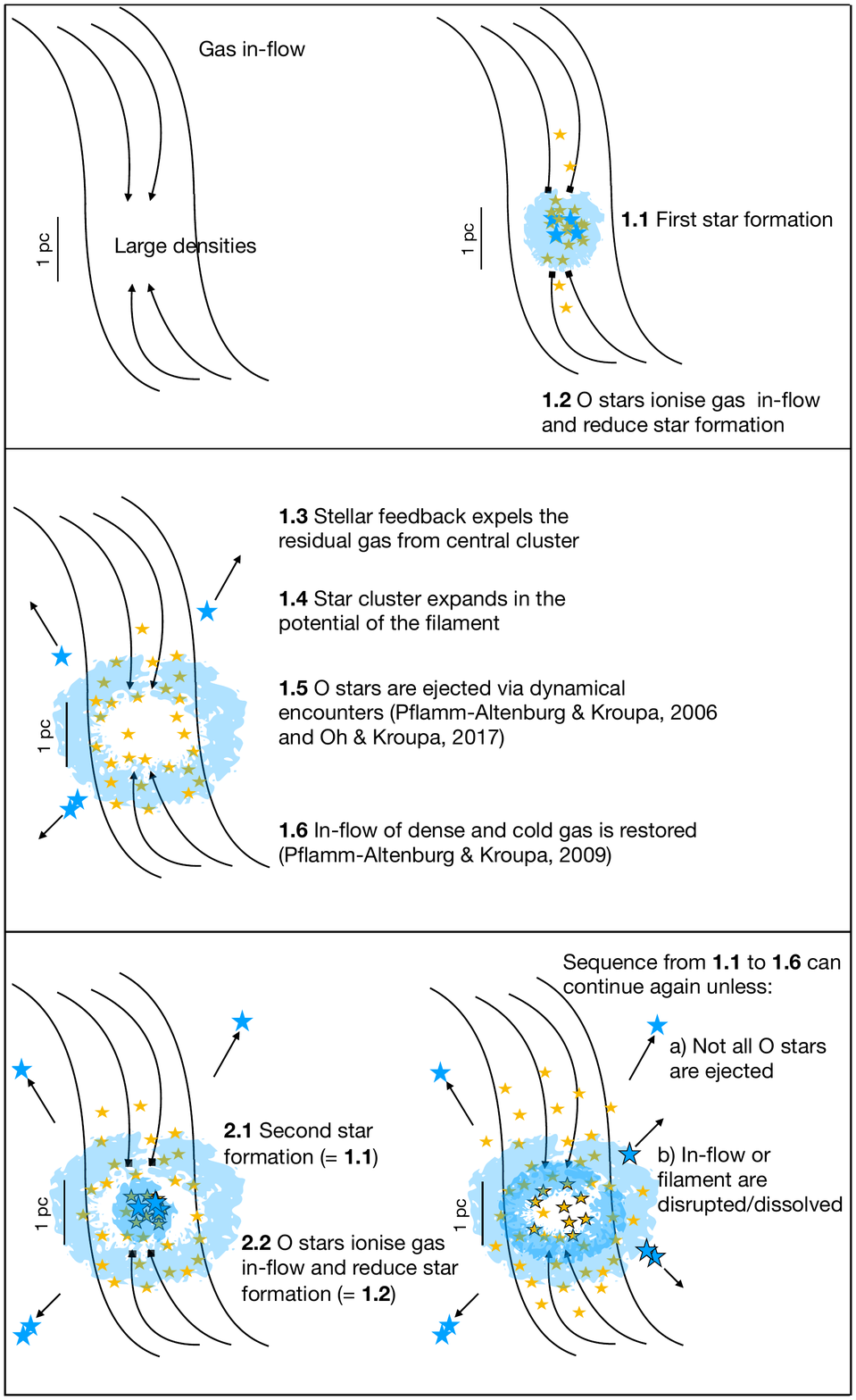}
		\caption{A schematic visualising the evolution of the molecular cloud
			filament, shown by the black curves, through the process of inflow
			(upper left panel), the monolithic formation of the first embedded
			cluster and termination of the inflow (upper right panel), the
			expansion of the first population due to gas expulsion and the
			ejection of the OB stars and resumption of gas inflow (middle
			panel), the monolithic formation of the
			second embedded cluster and termination of the inflow (lower left
			panel), the expansion of the second population due to gas expulsion
			and the ejection of its OB stars and resumption of gas inflow (lower
			right panel).
			\label{fig:steps}}
	\end{center}
\end{figure*}

A sketch of the here presented scenario is shown in
Fig.~\ref{fig:steps}.  Next we estimate the masses of the formed
populations and the number of ionising stars these are likely to have
had.

\subsection{The three populations}
\label{sec:3pops}

For a given mass of a stellar population one can estimate the masses
of the ionising (O, B) stars, once the IMF and sampling method
(random, optimal) is assumed. Since the past inflow history is not
known, the masses of the populations are estimated from the observed
relative population numbers as determined by \cite{Beccari+17} and
assuming a total mass of the ONC stellar population. Given an estimate
of the population's mass, the IMF allows a quantification of the
number of ionising stars.  The IMF in nearby star forming regions can
be well described by an invariant canonical IMF \citep{Kroupa01,
  Kroupa02b, Bastian10}. The canonical IMF can be conveniently
represented by two power-law segments,
$dN/dm \propto m^{-\alpha_{1,2}}$, where $\alpha_1=1.3$ for
$0.08\le m/M_\odot < 0.5$ and $\alpha_2=2.3$ (the Salpeter value) for
$0.5 \le m/M_\odot < m_{\rm max}$. Here, $dN$ is the infinitesimal
number of stars in the infinitesimal stellar-mass interval $m, m+dm$,
and $m_{\rm max}$ is the most massive star in the population. To
sample stellar populations, two approaches are employed here:
\begin{enumerate}

\item Mass-constrained random sampling: the IMF is sampled randomly
  until the desired population mass is achieved within a tolerance of
  $0.09\,M_\odot$, rejecting stars until the population mass is
  reached with a deviation of up to this amount. Here
  $m_{\rm max}=150\,M_\odot$ is the fundamental upper mass limit
  \citep{WK04, Figer05, OC05, Koen06, Banerjee12}. Since random
  sampling gives Poisson scatter to the final stellar population, each
  of the three populations is sampled 10000 times and the mean values
  and values corresponding to $1-\sigma$ dispersions are listed in
  Table~\ref{tab:Ostarsr}.

\item Optimal sampling \citep{Kroupa13}: the most massive star is
  given by the total mass of the co-eval stellar population which has
  a distribution of stellar masses following the canonical IMF. This
  sampling method leads naturally to the observed
  $m_{\mathrm{max}}-M_{\mathrm{ecl}}$ relation, \citep{Weidner06,
    KM11, KM12, Weidner13, Kroupa13, Megeath16, Ramirez16,
    Stephens+17, Yan+17}. The mathematical procedure of optimal
  sampling may be interpret physically as reflecting perfect
  self-regulation of the forming stars and star cluster \citep{AF96,
    Kroupa13}. It has been suggested that the available data appear to
  prefer optimal over random sampling (for example \citealt{Yan+17}).
  From Fig.~1 in \cite{Yan+17} or Fig.~4-5 in \cite{Kroupa13} it
  follows that a canonical co-eval stellar population with a mass of
  $M_{\rm ecl}\approx 320\,M_\odot$ contains one most-massive star
  with $m_{\rm max}\approx 19\,M_\odot$. Although $8\,M_\odot$ stars
  are also sufficiently luminous to ionise molecular gas, here a
  conservative approach is taken by considering only stars with
  spectral types earlier than about O8.9 as being particularly
  destructive.  Stellar-dynamical ejections of such stars become
  effective for canonical co-eval stellar populations with
  $320 < M_{\rm ecl}/M_\odot < 2000$ containing from a few OB stars to
  a few~O stars, respectively, noting that~B star binaries can be
  effective in ejecting an O~star. To sample stars optimally the
  GalIMF module from \cite{Yan+17} is employed\footnote{The source
    code is freely available on the github repository.  The GalIMF
    version 1.0.0 with associated programs is available at
    https://github.com/Azeret/galIMF, which is the whole galIMF
    branch. The here-used code is found in the example\_OS\_EC.py
    file.}.

\end{enumerate}
For all computations it is assumed that the total existing
stellar mass of the ONC today is
$M_{\mathrm{ONC}}\approx 1500\, M_{\odot}$, with the formation of the
third stellar population still continuing.  This allows an estimation
of the stellar masses of the individual existing stellar populations
(Table~\ref{tab:Ostarsr}), given that \cite{Beccari+17} provide
measurements of their relative fractions.

The mass of the third and currently forming population is calculated
assuming that $\theta^1$C~Ori is a binary with a total mass of
$44\pm7\,M_\odot$ \citep{Kraus+09} such that the most-massive and
second-most massive star in the population has a combined mass near
this value.  This gives a model mass of the third population of
$650\,M_\odot$, which then allows the calculation of the individual
stellar masses listed in Table~\ref{tab:OSpop}, using optimal
sampling.

It is to be noted that the sequences of masses given here depends on
the population masses and should therefore be taken to be illustrative
rather than definitive. Nevertheless, the masses of individual stars
are in agreement with the observational data. For example, the three
most massive stars in the first (oldest) population appear to be
similar to the three runaway O~stars ejected from the ONC
about~2.5~Myr ago (see Sec.~\ref{sec:disc}). By construction, the
first and second most massive star of the third (now forming)
population combined ($45.6\,M_\odot$), correspond very closely to the
mass of the binary system $\theta^1\,$C~Ori \citep{Kraus+09}, this
being a result of the model mass estimate used in optimal sampling.
The number of stars in the third population more massive than
$5\,M_\odot$ is in agreement between model and the observed ONC, which
currently has~ten of these \citep{PAK06}. We note that \cite{PAK06}
assumed a population mass of $1600\,M_\odot$ for the ONC. It must
therefore have ejected 30 stars more massive than $5\,M_\odot$, which
is similar to what the present model also implies (see below).

\begin{table}[t] \centering \begin{tabular}{c c || c | c }
		& & $\# \star: m_{\star}>19 M_{\odot}$ & $\# \star: m_{\star}>5 M_{\odot}$   \\ \hline
\textcolor{blue}{$1^{\mathrm{st}}$ pop.} & RS & $3 \pm 2$ & $18 \pm 3$ \\
 \textcolor{blue}{$886 M_{\odot}$}       & OS & 2 & 18  \\ \hline
\textcolor{yellow!80!red}{$2^{\mathrm{nd}}$ pop.} & RS &$2 \pm 1$ & $10 \pm 3$ \\
\textcolor{yellow!80!red}{$486 M_{\odot}$} & OS & 1 &  10 \\ \hline
\textcolor{red}{$3^{\mathrm{th}}$ pop.}&  RS & $2 \pm 1$ & $13\pm 3$\\
\textcolor{red}{$650 M_{\odot}$} & OS & 1 & 13 \\
                            \end{tabular} \centering \caption{Estimated mass in each of
                              the three populations (left column), the associated number of O stars
                              (central column) and of the number of all stars more massive than $5\,M_\odot$
                              (right column).  ``RS'' and ``OS'' refers to random sampling
                              and optimal sampling of the IMF, respectively.}
	\label{tab:Ostarsr}\end{table}

\begin{table}[t] \centering \begin{tabular}{c | c | c }
	$1^{\mathrm{st}}$, $886 M_{\odot}$	& $2^{\mathrm{nd}}$,  $486 M_{\odot}$ & $3^{\mathrm{rd}}$, $650 M_{\odot}$  \\ \hline \hline
[$M_{\odot}$] & [$M_{\odot}$]	&[$M_{\odot}$]\\	\hline
32.9 & 22.3 & 27.0 \\
22.9 & 15.3 & 18.6 \\
18.0 & 11.8 & 14.5 \\
15.0 & 9.8 & 12.0 \\
13.0 & 8.4 & 10.4 \\
11.4 & 7.4 & 9.1 \\
10.2 & 6.7 & 8.2 \\
9.3 & 6.1 & 7.5 \\
8.6 & 5.6 &  6.9 \\
8.0 & 5.2 & 6.4 \\
7.4 &  -- & 5.9 \\
7.0 &  -- & 5.6 \\
6.6 & -- & 5.3 \\
6.2 & -- &--  \\
5.9 & -- &-- \\
5.7 & -- &--  \\
5.4 & -- &--  \\
5.2 & -- &--  \\
                            \end{tabular} \centering \caption{The
                              exact model stellar populations (over $5\,M_\odot$)
                              assuming optimal sampling, a canonical IMF and the masses of
                              the three populations as described in the text. 
                            }
	\label{tab:OSpop}
\end{table}

\subsection{Suppressing star formation}
\label{sec:halt}

Having specified the three stellar populations, the question addressed
now is whether the populations can ionise the inner region of the
embedded cluster while not destroying the entire cloud within more
than a pc radius, before the ionising stars eject each other out of
the forming embedded cluster.

Once formed, massive stars ($m > 8 \,M_\odot$) ionise the surrounding
gas almost instantly, heating it up from $10-100 \,$K to
$\simeq 10^4 \,$K.  In a static homogeneous neutral cloud, the radius
within which the gas is ionised is the Str\"omgren radius,
\begin{equation}
R_{\mathrm{S}} = \left( \frac{3 \gamma_{\rm ion}}{4 \pi \beta^{*} n_{\rm H}^2} \right)^{1/3},
\label{eq:estroemgren}
\end{equation}
where $\beta^{*}=2.59\times 10^{-13}\,{\rm cm}^3 {\rm s}^{-1}$
\citep{Draine2011} is the hydrogen recombination coefficient into
excited states and $\gamma_{\rm ion}$ is the number of ionising
photons/s produced by the stellar population. This radius is given by
the balance between ionisation and recombination.  The pressure in the
ionised region is increased by approximately~3 orders of magnitude so
that the ionised region tends to expand into the cold cloud.  The
ionised medium inside the HII region is separated from molecular or
atomic gas outside by an ionisation front.  For a source producing
$\gamma_{\rm ion}$ ionising photons per second placed in a homogeneous
medium of particle density $n_{\rm H}$, and neglecting the
self-gravity, the position $R_{\rm IF}$ of the ionisation front
expands with time $t$ as (cf. \citealt{Spitzer1978, Hosokawa2006})
\begin{equation}
R_{\mathrm{IF}} = R_{\mathrm{S}} \left(1 + \frac{7}{4} \sqrt{\frac{4}{3}} \frac{c_s t}{R_{\mathrm{S}}} \right)^{4/7},
\label{eq:eexpansion}
\end{equation}
where $c_s\approx 10\,$km/s is the sound speed in the HII region.

Comparing the number of observed ultracompact (UC)HII regions (size
$< 0.1 \,$pc) with the number of observed O stars, and taking into
account the typical life-time of O stars, \cite{Wood1989} infer that
UCHII regions last for significantly longer ($\approx 0.1 \,$Myr) than
expected from Eq.~\ref{eq:eexpansion}.  Instead of expansion, some
UCHII regions show inward motions \citep{Wood2006, Klaassen2007}.

The observed longevity and inward motions of UCHII regions can be
explained by accretion at high rates ($\approx 10^{-3} \,M_\odot/$yr)
towards the central parts of the forming cluster as shown by detailed
3D hydrodynamical simulations performed by \cite{Peters2010a,
  Peters2010b}.  The gravitational well near massive stars is so deep
that accretion flows are accelerated to a velocity enough to swamp the
immediate surroundings of the young massive stars so that the UCHII
region cannot continuously expand, and Eq.~\ref{eq:eexpansion} cannot
be applied.  However, because the inner regions infall faster than the
outer regions, the influx of matter decreases, and the accretion disk
\citep{Chini+04, Chini+06, Nuernberger+07} around the forming
proto-star allows the polar regions to be ionised \citep{Nielbock+07}.
The UCHII region is highly non-spherical, and the ionisation front
rapidly fluctuates.  \citet{DePree2014} observe fluctuations or
flickering of 41~UCHII regions in the Sgr B2 region, and find very
good agreement with the explanation proposed by \citet{Peters2010a}.
About~95~\% of the mass of the final star accretes within
about~0.1~Myr \citep{WT03, Duarte-Cabral+13}, setting the life-times
of the UCHII regions.

After the initial UCHII stage, massive stars clear the young cluster
from the remaining gas: Observations \citep{LL03, Lada10, Andre+14,
  Megeath16} suggest that more than 60\% of the total mass of embedded
star clusters is in the form of gas being removed by the action of
massive stars.  Another piece of evidence for rapid (at velocity
$\approx 10 \,$km/s) removal of $\approx2/3$ of the total mass in the
form of gas comes from the comparison between the observed density and
velocity profiles of very young clusters (ONC, NGC~3603, R~136) and
Nbody simulations \citep{BK15b}.  Rapid removal of this amount of mass
can also explain the increase of radii from the embedded stage
(comparable to the widths of the molecular-gas filaments,
\citealt{MK12}) to the current stage of exposed star clusters
\citep{BK17}.  High-resolution magnetohydrodynamical simulations of
proto-stellar formation also imply that a large fraction of the
accreting gas is channeled outwards into outflows \citep{MM12, Bate14,
  Federrath+14, Kuruwita+17}.

Another assumption which limits the validity of
Eq.~\ref{eq:eexpansion} is the homogeneity of the surrounding gas.
\cite{Whitworth1979} and \cite{Tenorio-Tagle1979} show that the HII
region expands further in the direction of the lowest column density
as observed from the ionising star.  When the gas reaches low density
regions of the molecular cloud, it erupts outwards forming a champagne
flow \citep{Tenorio-Tagle1979}.  When the eruption occurs, the HII
region is evacuated and its ability to shield the ambient cloud from
ionising photons decreases substantially. This happens because the
ionised hydrogen, which recombines on a time-scale of about 100~yr for
$n\approx 10^3\,$protons/cm$^3$, flows out, thus allowing more
ionising photons to reach the ionising front.  The cloud is rapidly
photo-evaporated from the inside, with the ionised gas streaming in
the opposite direction out of the cluster.  Currently, the HII region
ionised by the ONC is highly non-spherical, elongated towards the Sun,
with the Trapezium located near the side of the more distant edge of
the cloud \citep{Wen1995, Odell2001}.  Gas is photo-evaporated mainly
at this edge of the cloud streaming towards us.

\begin{figure}[ht!] \begin{center}
		\scalebox{1.0}{\includegraphics{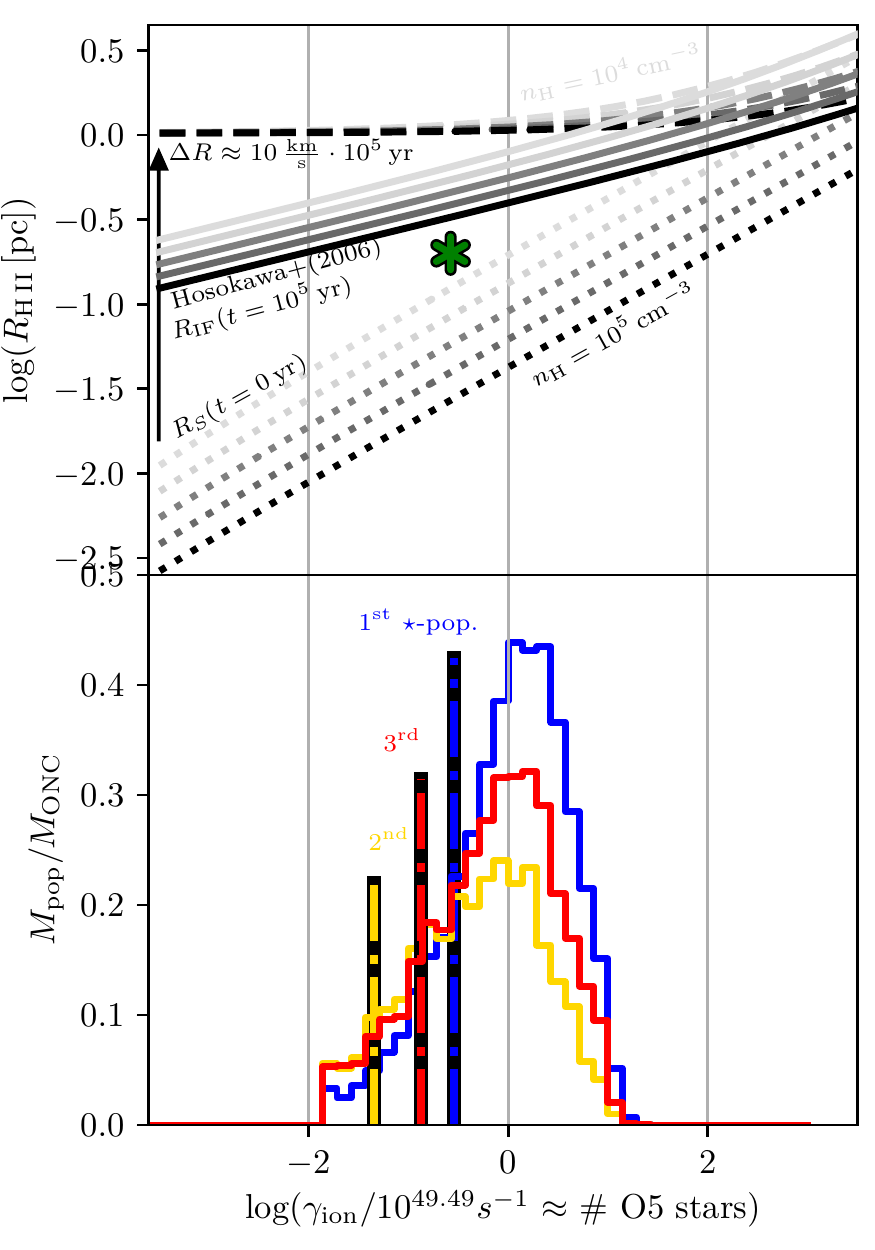}} \end{center}
              \caption{Distance, $R_{\mathrm{H\,II}}$, to the radial
                molecular cloud filament within which the gas is
                ionised in dependence of the number of ionising
                photons, $\gamma_{\rm ion}$, whereby one~O5V star
                emits $\gamma_{\rm ion}=10^{49.49}\,$ionising
                photons/s according to \cite{Sternberg2003}. {\bf
                  Bottom panel:} The histograms show the range of
                values obtained for the random sampling case (10000
                realisations), and the thick vertical lines refer to
                the optimal sampling cases which have no Poisson
                spread. The ionising flux from all stars in a co-eval
                population is summed to obtain $\gamma_{\rm ion}$.
                The maximum value of the histogram is proportional to
                the mass of the population.  The color code is the
                same as in Table~\ref{tab:Ostarsr}. Thin gray vertical
                lines are presented for better orientation in the
                figure. {\bf Top panel:} The dotted lines are
                $R_{\rm S}$, which are the initial radii of the UCHII
                regions (Eq.~\ref{eq:estroemgren}) for different
                values of the density of the cloud, $n_{\rm H}$,
                assuming no change in density within the UCHII region
                compared to the surrounding molecular cloud. The solid
                lines are expanded HII regions
                (Eq.~\ref{eq:eexpansion}) for the same set of
                $n_{\rm H}$ values as for the dotted lines.  The
                dashed lines indicate $R_{\rm HII}$ assuming the
                density of the ionised gas is zero up until the
                ionisation front. This front expands with $10\,$km/s
                such that at $10^5\,$yr $R_{\rm HII}\approx 1\,$pc for
                $\gamma_{\rm ion} < 10^{49.49}\,$photons/s, being
                progressively larger for larger photon fluxes.  For
                example, the first (oldest, blue) population, if it by
                chance were to contain 10~O5V stars, would have
                produced an UCHII region with
                $R_{\rm HII}\approx 10^{-1}\,$pc. If this UCHII region
                can break out as a champagner flow, it expands within
                about~0.1~Myr to $R_{\rm HII}\approx 1\,$pc (note that
                $10^5\,$cm$^{-3} = 2500\,M_\odot$pc$^{-3}$ is about
                the best vacuum achievable on Earth).  The present-day
                size of the observed HII region in the ONC, the green
                asterisk \citep{Felli1993}, is consistent with our
                estimates.  } \label{fig:ion}
\end{figure}

Given the complexity of the configuration mentioned above, the
detailed damage done to the cloud by the ionising stars need to be
accessed with detailed radiation-transfer-hydro-dynamic simulations.
In this work order-of-magnitude estimates are performed which is
sufficient to study if the here proposed mechanism for producing
multiple co-eval populations in a VYC may be plausible. To do so, the
ionising fluxes of all stars of one co-eval population are combined to
calculate $\gamma_{\rm ion}$ using the data provided by
\cite{Sternberg2003}.  The upper limit of the extent of the ionised
region is calculated assuming that the champagne flow occurs
immediately and that the ionisation front advances inwards the cloud
and away from the co-eval population with a velocity of $10 \,$km/s
for $0.1 \,$Myr.  The initial radius of the ionisation front is
assumed to start at $R_{\rm S}$ for a plausible range of particle
densities for the ONC \citep{Felli1993}.  The estimates of the
distance from the source of the ionisation front are evident in the
upper panel of Fig.~\ref{fig:ion}. The abscissa shows the ionising
photon production rate, $\gamma_{\rm ion}$, in units of O5V stars (the
luminosity of an O5V star is adopted from table~1 of
\citealt{Sternberg2003}).  The distance from the source to the
ionisation front is shown by dashed lines in the upper panel of
Fig.~\ref{fig:ion}.  The Str\"{o}mgren radius is plotted for the
plausible density range (see Eq.~\ref{eq:estroemgren}), representing
the lower estimate by dotted lines.  The estimate based on
Eq.~\ref{eq:eexpansion} is plotted by solid lines.  The current state
of the HII region in the ONC, as shown by the green asterisk based on
the date from \cite{Felli1993}, is in agreement with the estimates.
Assuming that massive stars are present for a few~$0.1 \,$Myr, these
estimates indicate that massive stars from the previous two
populations were able to prevent star formation and largely dispersed
the innermost $0.1 - 1.0 \,$pc of the filamentary cloud, but more
distant parts were preserved and could resume falling inwards into the
cluster.

\subsection{Competition of time-scales}
\label{sec:comp}

In general, for a sufficiently low-mass proto-cluster of stars, which
contains only a few hundred stars and thus at most a few OB stars, the
central core containing the OB stars which are almost exclusively
formed in energetic binaries \citep{Sana+12,MDiS17}, is dynamically
very unstable.  Ejection of the massive stars occurs through energetic
three- and four-body encounters in the core, and is significantly
enhanced for a realistic binary population. This process of producing
runaway massive stars into the Galactic field has been calculated for
a large ensemble of young clusters using the Aarseth-Nbody6 code by
\cite{OKP15}.  The result is that the observed runaway and isolated O
star population in the Galactic field can be well accounted for by
realistic clusters and binary populations.  The core ejects its
massive members within a few crossing-time scales, that is within
$<10^5\,$yr \citep{PAK06}. A large theoretical survey of the dynamical
evolution of young clusters containing all stars initially in a
realistic binary population \citep{Sana+12,Belloni+17} shows that
clusters with a mass near $10^{2.5}$ and $10^3\,M_\odot$ can expel all
their few most massive stars \citep{OKP15, OK16}.

How likely is it that all massive stars are ejected from the low-mass
embedded cluster within~1~Myr? \cite{PAK06} performed a survey using
high accuracy and -precision chain-regularisation dynamical modelling
of Trapezium-like clusters containing 4, 10 and 40 stars more massive
than $5\,M_\odot$ and reflecting the masses of the observed ten
massive stars in the ONC. The result is that all such stars are
ejected in about 45~(62)~per cent of the cases after 0.5~(1.0)~Myr for
clusters with initially~4 massive stars, all stars are ejected in
about 32~(45)~per cent of the cases after 0.5~(1.0)~Myr for clusters
with initially 10 massive stars, and all stars are ejected in about
1~(7)~per cent of the cases after 0.5~(1.0)~Myr for clusters with
initially 40 massive stars.  The ejection of all most-massive stars is
thus not rare for low-mass $<10^3\,M_\odot$ embedded clusters. If one
considers only stars more massive than $19\,M_\odot$, then the cases
when all such stars are ejected would be more frequent, extending the
range of masses of co-eval populations to $320-2000\,M_\odot$
(Sec.~\ref{sec:3pops}).  Clusters which do not eject all their
ionising stars will not be able to form multiple co-eval populations
distinctly separated by about~1~Myr.

A competition in time-scales therefore exists: On the time-scale of a
few 0.1~Myr the population builds up from the inflow of molecular
gas. When the massive star(s) form, the inner~0.1~pc region is ionised
and the UCHII region breaks out after a time-scale of about
0.1~Myr. On the same time-scale, the massive stars eject each other
from the population, as shown for such small system to be the case by
direct Nbody simulations \citep{PAK06}. The molecular inflow can
resume, forming a second population, separated in time from the
previous one by about~1~Myr. The separation in time is uncertain, but,
given that embedded populations need a few~0.1~Myr to form, the
ejections of the massive stars need~1~Myr (longer if the massive stars
need to mass-segregate to the centre, Fig.~7 in \citealt{OK16}), a
separation by about 0.5--1~Myr appears physically
plausible. Conversely, by observing the separation time scale, as is
the case for the data obtained by \cite{Beccari+17}, it now appears
possible to map out the sequence of events that occurred in the ONC.

\subsection{A possible history of mass accretion, and the IMFs}
\label{sec:hist}

\begin{figure}[ht!] \begin{center}
		\scalebox{1.0}{\includegraphics{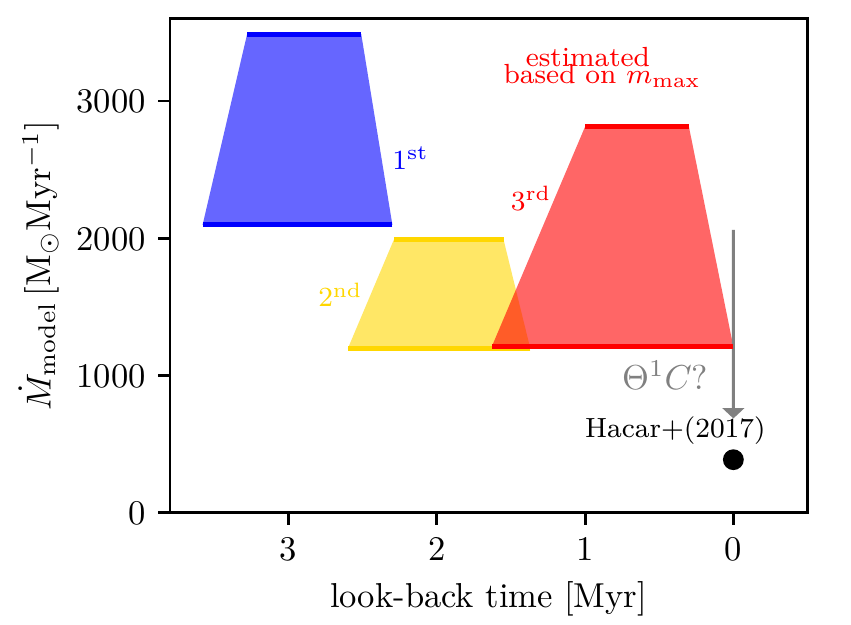}} \end{center}
              \caption{The estimated mass inflow rates,
                $\dot{M}_{\rm model}$, which may have occurred
                during the successive build-up of the stellar
                population of the ONC. Inflow began about 3~Myr ago
                when the molecular-cloud filament, at present in the
                shape of an integral, established itself.  This formed
                the first (blue) population. The presently on-going
                inflow, shown in red, has been forming the third
                (present-day) population and is larger than the
                directly measured value by \cite{Hacar+17}. This may
                be a sign of on-going photo-ionisation of the inflow
                through the star $\theta^1$C~Ori. The upper and lower
                bounds on $\dot{M}_{\rm g, flow}$ are the one-sigma
                and the 5--95~per cent bounds, respectively, on the
                time-scales over which the populations formed (table~1
                in \citealt{Beccari+17}).  } \label{fig:mdot}
\end{figure}

In an attempt to re-create the inflow history into the ONC, the masses
of the three populations are corrected for a SFE of 33~per cent and
divided by the one-sigma and 5--95~per cent age time-intervals for
each of the first, second and third population (table~1 in
\citealt{Beccari+17}). For the presently forming one it is assumed it
formed over a time interval of $1.04\,$Myr (lower limit) or 1.63~Myr
(upper limit on the age from \citealt{Beccari+17}). The so estimated
mass-inflow rates are compared to the currently observed value in
Fig.~\ref{fig:mdot}.

\begin{figure*}[ht!] \begin{center}
		\scalebox{1.0}{\includegraphics{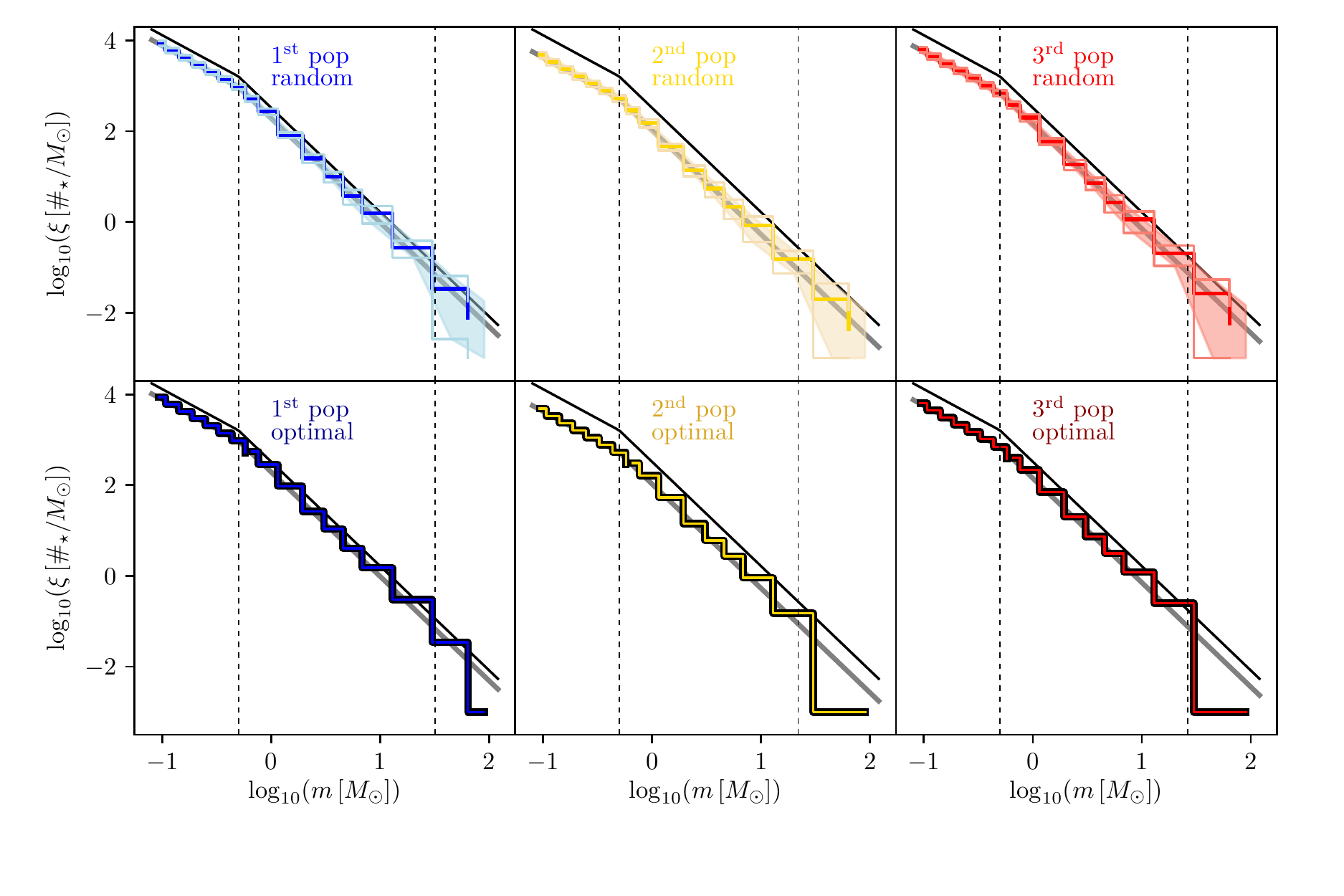}} \end{center}
              \caption{The predicted model IMFs of the three
                populations, colour coded as in
                Table~\ref{tab:Ostarsr}. The upper row shows the IMFs
                assuming random sampling. The shaded regions indicate
                the one-sigma range for 10000 realisations.  The lower
                row shows the IMFs assuming nature follows optimal
                sampling. These have no uncertainties. The solid lines
                indicate the canonical IMF normalised to the empirical
                total stellar mass of all populations combined as
                given by \cite{Beccari+17}, and the right dashed
                vertical lines depict $m_{\rm max}$ while the left
                vertical lines indicate $0.5\,M_\odot$. The thick grey
                solid lines are the canonical IMF normalised, as are
                the histograms, to the mass of each
                sub-population. Note that if the IMF is a probability
                distribution function (upper panels) then the distinct
                co-eval populations would on average contain a larger
                number of more-massive stars than if the IMF is an
                optimally sampled distribution function (lower
                panels).  } \label{fig:IMF}
\end{figure*}

Given that \cite{Beccari+17} has now the means to separate the
populations in the ONC in the colour-magnitude diagramme, it has
become possible to observationally estimate their individual IMFs, by
performing the complex transformation of luminosities to stellar
masses for PMSs, taking into account that a large fraction of these is
in unresolved multiple systems \citep{KTG91, Kroupa13}. Within the
present model it is possible to calculate the model IMFs of the three
populations as a prediction, as shown in Fig.~\ref{fig:IMF}.

\section{Discussion}
\label{sec:disc}

Proto-clusters with a molecular-gas inflow rate of
$\dot{M}_{\rm g, flow}=1000-6000\,M_\odot\,{\rm Myr}^{-1}$ over a
time-scale of about $1\,$Myr can form a co-eval population of stars
with a mass in the range $M_{\rm ecl}=300-2000\,M_\odot$. About
40--60~per cent of these will be able to expel all their O stars
within $\approx 10^6\,$yr after the formation of the O stars
\citep{PAK06}.  Since such co-eval populations have too few O~stars to
disperse the whole molecular cloud over scales larger than a~pc, such
systems are likely to be able to resume the inflow of molecular gas
\citep{PAK09}, and to form a second co-eval population with an age of
about $1\,$Myr older than the first, depending on how quickly the
inflow resumes and begins to form a new population. In about 60~per
cent of such cases the same process may repeat. The ONC, with
currently three populations separated in age by about $1\,$Myr,
may have undergone such a history of {\it repeated stellar-dynamical
  termination of feedback-halted filament-accretion}. The ONC,
observed at the present day, may thus be in the process of halting the
formation of its third generation of stars with the Trapezium at its
centre being in the process of decaying stellar-dynamically (see
below).

This scenario appears to be plausible as it combines
reasonably-understood physical processes which are the destructive
photo-ionising effect of O stars on molecular clouds and
stellar-dynamical ejections. It carries a number of predictions:

Firstly, clusters less massive than about $300\,M_\odot$ should not
show multiple populations of stars separated in age with age gaps less
than a few~Myr since in such clusters O~stars would not form (for
example there are no O stars in the whole combined population of in
total a few~$10^3$ stars formed in many small clusters in the southern
part of the Orion~A cloud, \citealt{Hsu12, Hsu13}).  During cluster
assembly, the inflow-rate would be smaller than about
$1000\,M_\odot\,$Myr$^{-1}$, such that an embedded cluster with a
stellar mass $M_{\rm ecl}<300\,M_\odot$ may form within about 1~Myr.
Formation ceases when the inflow decreases given a lack of gas supply,
for example because the molecular cloud disperses.
Magneto-hydrodynamic simulations show that about 50~per cent of the
gas is ejected by the outflows alone \citep{MM12, Federrath+14,
  Bate+14}. In combination with other feedback such as radiation
feedback, the SFE may reach 33~per cent or less within the
embedded-cluster volume of typically less than $1\,$pc$^3$
\citep{MK12}.  This is indeed shown to be the case by observational
constraints \citep{Andre+14, Megeath16}.

Secondly, clusters which form the first co-eval population more
massive than $M_{\rm ecl} \approx 2\times 10^3\,M_\odot$ are never
likely to expel all their O stars \citep{OKP15}.  The inflow of the
molecular gas will be terminated and cannot resume because of the
photo-destruction of the molecular cloud. For a SFE of 33~per cent and
a time-scale of formation of 1~Myr, this yields a mass-inflow rate
$6 \times 10^3\,M_\odot$Myr$^{-1}$.  Such and more massive VYCs are
therefore not likely to show the fine stratification of multi-age
populations evident in the ONC, but such clusters may be contaminated
by captured field stars \citep{PAK07}.  Also, such clusters may
accrete gas from a molecular cloud once the precursor stars of
core-collapse supernovae ($m>8\,M_\odot$) have died, thus allowing
populations of stars separated in age by a few dozens to 50~Myr
\citep{PAK09, BJK17}.  That gas-accretion may indeed be leading to the
formation of new stars in such clusters has been suggested by
observation \citep{FB17}.

Thirdly, it follows that if the above process is active then an
embedded cluster can build-up a population of stars multiple
times. For example, if three co-eval populations can form then the
range of masses of the combined stellar population of the VYC can be
$900-6000\,M_\odot$. If the process repeats on a time-scale of a Myr
for the life-time of a molecular cloud, 10~Myr, then the VYC could
reach a combined stellar mass of $3000-2\times 10^4\,M_\odot$.
Therefore, VYCs with combined populations in the mass range
$300< M_{\rm ecl}/M_\odot<2\times 10^4$ may show a fine age
stratification with populations separated by $\approx1\,$My. A rough
estimate based on the available work \citep{PAK06, OKP15} suggests
that about 20~per cent of all such clusters ought to show three age
sequences (assuming there is a chance of 60~per cent that each expels
all its ionising stars within 1~Myr). To better ascertain how frequent
this multi-population occurrence is, we need further numerical
simulations using highly-precise and accurate Nbody integrators. It is
plausible that the vast majority of embedded clusters containing two
to three~O star binaries will eject all O stars, in this event making
this multi-age phenomenon the rule rather than the exception.

Fourthly, according to the present model the ONC should have ejected,
in total, about 31 stars more massive than $5\,M_\odot$
(Table~\ref{tab:Ostarsr}), given that currently 10 are observed to
still be within it \citep{PAK06}. Since the four stars AE~Aur,
$\mu$~Col and $\iota$~Ori (binary, see below) have already been
ejected, there should now be about~27 stars more massive than
$5\,M_\odot$ moving away from the ONC. About 44~per cent of these
should have been ejected about 2.9~Myr ago, 24~per cent about 1.9~Myr
ago, and 32~per cent are being ejected now (using the optimal sampling
data in Tab.~\ref{tab:Ostarsr} in conjunction with the ages given in
tab.~1 in \citealt{Beccari+17}).  These should be found using Gaia
astrometry mission data. It is to be cautioned though that these are
rough estimates, based on the available observed population numbers by
\cite{Beccari+17} and some of the stars can be very far away (250~pc
for an ejection velocity of 100~km/s with ejection 2.5~Myr ago).
Also, the two-step ejection process \citep{Pflamm10} implies that some
of the ejected stars cannot be traced back to their birth sites. That
stars are being ejected is indeed observed \citep{Poveda+05, ChT12}.

The scenario discussed here may help to understand a number of
problems concerning the ONC:

Firstly, it has been noted that the proplyds (PMSs with their
accretion cacoons) in the ONC may have short ($<10^5\,$yr)
photo-evaporation time-scales posing a problem given that it has been
thought that the Trapezium is much older
\citep{ODell98}. \cite{ODell+09} discuss this issue at length and in
view of the complex gas flows within the ONC (see also
\citealt{McLeod+16}), suggesting that the present configuration may be
less than $10^5\,$yr old. The present model may thus resolve this
issue in that the inner ONC and the Trapezium may indeed be very young
and of the order of $10^5\,$yr old or even younger.

Secondly, it has been noted that in the inner region of the ONC the
stellar velocity dispersion is too large given the observed mass. One
interpretation is that the very recent expulsion of residual gas
associated with the presently forming generation of stars has lead to
the expansion of the population \citep{Kroupa+99, K2000, KAH,
  Subr+12}.  The Trapezium of massive stars at the core of the ONC is
also observed to have a too large velocity dispersion for its mass. It
may be possible that the Trapezium is at the moment undergoing
stellar-dynamical decay which would be consistent with the present
model. The large velocity of the main ionising star $\theta^1$C~Ori
may be a result of this decay, as discussed by \cite{ODell+09}.  It
has also been suggested that the Trapezium is bounded by the presence
of a black hole of mass near $100\,M_\odot$, which may have formed
from runaway merging and subsequent implosion of the merged massive
star \citep{Subr+12}). The present scenario may relax the rational for
this exotic, albeit interesting, possibility.

Thirdly, the {\it repeated stellar-dynamical termination of
  feedback-halted filament-accretion} model would explain why the
oldest stars with an age of about $3\,$Myr are most extended
surrounding also the integral-shaped filament over a spatial scale of
5--10~pc (see Fig.~\ref{fig:steps}). The first generation of stars
most likely expanded after the expulsion of residual gas and these
stars are now orbiting the general potential made by the ONC and the
integral-shaped filament. An embedded cluster which undergoes
realistic expulsion of residual gas expands significantly until a part
of it re-virialises at a half-mass radius about 3- to 5-times larger
than the half-mass radius of the pre-expulsion embedded cluster
\citep{KAH, Brinkmann+17}. However, such calculations also show that
the 50~per cent Lagrange radius of a cluster has expanded by a factor
of about ten after about 2.5~Myr (Fig.~1 in \citealt{KAH}). The oldest
ONC stars can thus be found at large distances around the Trapezium of
up to 5~pc or more.  The next generation of $1.9\,$Myr old stars is in
total a less massive population according to \cite{Beccari+17}, and
has therefore expanded less given the smaller velocity dispersion of
the pre-gas-expulsion embedded cluster which is embedded within the
expanded first population. The currently forming third population has
not yet reacted significantly to the most recent and on-going gas
blow-out and is therefore most-concentrated around the Trapezium. That
gas is currently being expelled from the ONC is suggested from the
rapid photo-erosion of the molecular material within the inner~pc
region centred on the ONC \citep{ODell+09}.

Fourthly, calculations of the trajectories of the stars AE~Aur
(spectral type O9.5V), $\mu$~Col (O9.5V) and $\iota$~Ori (O9III+B1III,
binary) by \cite{Hoogerwerf+01, Gualandris+04} indicate that they all
originated about~2.5 million years ago from a position close to the
ONC, supporting the present scenario (see also the discussion in
\citealt{ODell+09}). Note that, if the IMF is an optimally sampled
distribution function, then this first population will have had just
such stars as the most massive ones (Table~\ref{tab:OSpop}).  The
evident association of these stars with the ONC posed a long-standing
problem since it was thought that the ONC is younger than~2.5~Myr,
while if it had formed~2.5~Myr ago, its association with a molecular
cloud would appear unlikely given the destructive effect of its~O
stars. The present scenario may solve this problem by the ONC building
up over time through a series of monolithic star bursts, the first one
of which occurred about~3~Myr ago and which formed only a few massive
stars which ejected each other out of the cluster. AE~Aur, $\mu$~Col
and $\iota$~Ori may be just these stars.

\section{Conclusions}
\label{sec:concs}

If molecular cloud cores have inflow rates of molecular gas towards a
proto-cluster in the range $1000-6000\,M_\odot$/Myr then the
proto-cluster can buildup within~1~Myr a first co-eval population of
stars in the mass range $300-2000\,M_\odot$. Such a population is
likely to contain up to a few~O stars which are likely to eject each
other within~1~Myr. With the ionising sources gone, the inflow of
molecular gas can again reach the potential minimum of the
proto-cluster forming another similar population. Depending on how
long the inflow can be sustained, the final VYC can have a stellar
mass range of 1000 to $6000\,M_\odot$ for three such events for
example, the three populations being separated in age by about~1~Myr.
The mass range for VYCs to host multiple populations spans from
$600\,M_\odot$ (for two populations separated by one~Myr) to
$20000\,M_\odot$ (for ten populations, for a molecular cloud life time
of 10 Myr).

The ONC discussed here lies in this mass range, and NGC~2024 may be
another example at the lower end of this range \citep{Getman+14}.  The
physical process for creating such an age stratification is that the
inflowing molecular gas builds up a first generation of stars which
contain a few O stars. The O stars ionise the inflowing molecular gas
therewith inhibiting star formation within the embedded cluster,
although star formation may continue in the filament. These stars will
orbit through the ONC and will attain a larger spatial distribution.
In the majority of cases the O stars are ejected
stellar-dynamically. With the ionising sources gone, the ionised gas
can recombine and form molecules, and the still-inflowing gas can
reach the embedded cluster volume in molecular form and form a new
population of stars. This process can continue until the inflow tapers
out, or becomes so massive that too many massive stars form for all to
be ejected.  Fig.~\ref{fig:steps} shows a cartoon of this process. The
gas chemistry, which depends on the recombination and molecule
formation rates, may be an interesting problem to consider under these
conditions.

The model is here used to estimate the history of filamentary infall
of molecular gas into the ONC. The three O stars AE~Aur, $\mu$~Col and
$\iota$~Ori may be the stars ejected when the first ONC population
formed. The wider spatial dispersion of the older populations can be
understood as a result of the expulsion of residual gas due to the
feedback from the O stars, causing expansion of the respective
population, and as a result of the wider orbits of the stars formed in the
filament. When the O~stars are ejected and star-formation within the
inner ONC resumes, the whole population will contract somewhat due
to the deepened potential. The whole cycle would need to be simulated
self-consistently to gain more knowledge on the detailed dynamical and
star-formation history of each component. 

The here proposed repeated stellar-dynamical termination of
feedback-halted filament-accretion model needs to be studied in much
more detail, for example by performing gas-dynamical simulations with
star formation and radiative and mechanical feedback including the
self-consistent computation of high-precision and accurate
stellar-dynamical processes including chain-regularisation procedures
\citep{Aarseth03}, to ascertain if the broad scenario outlined here is
indeed viable.  The AMUSE software framework \citep{Portegies09} may
be interesting in this context.  At the same time, the accumulation of
data on the motions of individual stars in the various age groups, for
example with the GAIA mission, should lead to a better understanding
of the origin of the populations (for example as suggested in the
present model as being expanding post-gas-expulsion populations), or
if a very different model may be applicable. A different model based
on an oscillating filament which launches proto-stars has been
suggested by \cite{SG16, Boekholt+17} and may be relevant for
understanding the observed structure and motions within and around the
ONC.

\acknowledgements{We thank Monika Petr-Gotzens for useful comments. PK
  thanks the ESO office in Garching for supporting his stay there as a
  scientific visitor.  TJ was supported by Charles University in
  Prague through grant SVV-260441 and through a stipend from the
  SPODYR group at the University of Bonn.  FD acknowledges the support
  from the DFG Priority Program 1573 ``The physics of the interstellar
  medium''.  This project was initiated in September in Prag at the
  conference Modelling and Observing Dense Stellar Systems 2017
  (MODEST17). We thank, amongst the other organisers, Ladislav Subr
  and Jaroslav Haas for their organisational work.}

\bibliographystyle{aa}
\bibliography{kroupa_ref} 

\begin{thebibliography}{113}
\expandafter\ifx\csname natexlab\endcsname\relax\def\natexlab#1{#1}\fi

\bibitem[{{Aarseth}(2003)}]{Aarseth03}
{Aarseth}, S.~J. 2003, {Gravitational N-Body Simulations}

\bibitem[{{Adams} \& {Fatuzzo}(1996)}]{AF96}
{Adams}, F.~C. \& {Fatuzzo}, M. 1996, \apj, 464, 256

\bibitem[{{Andr{\'e}} {et~al.}(2014){Andr{\'e}}, {Di Francesco},
  {Ward-Thompson}, {Inutsuka}, {Pudritz}, \& {Pineda}}]{Andre+14}
{Andr{\'e}}, P., {Di Francesco}, J., {Ward-Thompson}, D., {et~al.} 2014,
  Protostars and Planets VI, 27

\bibitem[{{Banerjee} \& {Kroupa}(2013)}]{BK13}
{Banerjee}, S. \& {Kroupa}, P. 2013, \apj, 764, 29

\bibitem[{{Banerjee} \& {Kroupa}(2014)}]{BK14}
{Banerjee}, S. \& {Kroupa}, P. 2014, \apj, 787, 158

\bibitem[{{Banerjee} \& {Kroupa}(2015)}]{BK15}
{Banerjee}, S. \& {Kroupa}, P. 2015, \mnras, 447, 728

\bibitem[{{Banerjee} \& {Kroupa}(2017)}]{BK17}
{Banerjee}, S. \& {Kroupa}, P. 2017, \aap, 597, A28

\bibitem[{{Banerjee} \& {Kroupa}(2018)}]{BK15b}
{Banerjee}, S. \& {Kroupa}, P. 2018, in Astrophysics and Space Science Library,
  Vol. 424, The Birth of Star Clusters, ed. S.~{Stahler}, 143

\bibitem[{{Banerjee} {et~al.}(2012){Banerjee}, {Kroupa}, \& {Oh}}]{Banerjee12}
{Banerjee}, S., {Kroupa}, P., \& {Oh}, S. 2012, \mnras, 426, 1416

\bibitem[{{Bastian} {et~al.}(2010){Bastian}, {Covey}, \& {Meyer}}]{Bastian10}
{Bastian}, N., {Covey}, K.~R., \& {Meyer}, M.~R. 2010, \araa, 48, 339

\bibitem[{{Bate}(2014)}]{Bate14}
{Bate}, M.~R. 2014, \mnras, 442, 285

\bibitem[{{Bate} {et~al.}(2014){Bate}, {Tricco}, \& {Price}}]{Bate+14}
{Bate}, M.~R., {Tricco}, T.~S., \& {Price}, D.~J. 2014, \mnras, 437, 77

\bibitem[{{Beccari} {et~al.}(2017){Beccari}, {Petr-Gotzens}, {Boffin},
  {Romaniello}, {Fedele}, {Carraro}, {De Marchi}, {de Wit}, {Drew}, {Kalari},
  {Manara}, {Martin}, {Mieske}, {Panagia}, {Testi}, {Vink}, {Walsh}, \&
  {Wright}}]{Beccari+17}
{Beccari}, G., {Petr-Gotzens}, M.~G., {Boffin}, H.~M.~J., {et~al.} 2017, \aap,
  604, A22

\bibitem[{{Bekki} {et~al.}(2017){Bekki}, {Je{\v r}{\'a}bkov{\'a}}, \&
  {Kroupa}}]{BJK17}
{Bekki}, K., {Je{\v r}{\'a}bkov{\'a}}, T., \& {Kroupa}, P. 2017, \mnras, 471,
  2242

\bibitem[{{Belloni} {et~al.}(2017){Belloni}, {Askar}, {Giersz}, {Kroupa}, \&
  {Rocha-Pinto}}]{Belloni+17}
{Belloni}, D., {Askar}, A., {Giersz}, M., {Kroupa}, P., \& {Rocha-Pinto}, H.~J.
  2017, \mnras, 471, 2812

\bibitem[{{Boekholt} {et~al.}(2017){Boekholt}, {Stutz}, {Fellhauer},
  {Schleicher}, \& {Matus Carrillo}}]{Boekholt+17}
{Boekholt}, T.~C.~N., {Stutz}, A.~M., {Fellhauer}, M., {Schleicher}, D.~R.~G.,
  \& {Matus Carrillo}, D.~R. 2017, \mnras, 471, 3590

\bibitem[{{Bonnell} \& {Smith}(2011)}]{BS11}
{Bonnell}, I.~A. \& {Smith}, R.~J. 2011, in IAU Symposium, Vol. 270,
  Computational Star Formation, ed. J.~{Alves}, B.~G. {Elmegreen}, J.~M.
  {Girart}, \& V.~{Trimble}, 57--64

\bibitem[{{Bontemps} {et~al.}(2010){Bontemps}, {Motte}, {Csengeri}, \&
  {Schneider}}]{Bontemps+10}
{Bontemps}, S., {Motte}, F., {Csengeri}, T., \& {Schneider}, N. 2010, \aap,
  524, A18

\bibitem[{{Bressan} {et~al.}(2012){Bressan}, {Marigo}, {Girardi}, {Salasnich},
  {Dal Cero}, {Rubele}, \& {Nanni}}]{Bressan+12}
{Bressan}, A., {Marigo}, P., {Girardi}, L., {et~al.} 2012, \mnras, 427, 127

\bibitem[{{Brinkmann} {et~al.}(2017){Brinkmann}, {Banerjee}, {Motwani}, \&
  {Kroupa}}]{Brinkmann+17}
{Brinkmann}, N., {Banerjee}, S., {Motwani}, B., \& {Kroupa}, P. 2017, \aap,
  600, A49

\bibitem[{{Burkert}(2017)}]{Burkert17}
{Burkert}, A. 2017, ArXiv e-prints

\bibitem[{{Chatterjee} \& {Tan}(2012)}]{ChT12}
{Chatterjee}, S. \& {Tan}, J.~C. 2012, \apj, 754, 152

\bibitem[{{Chini} {et~al.}(2004){Chini}, {Hoffmeister}, {Kimeswenger},
  {Nielbock}, {N{\"u}rnberger}, {Schmidtobreick}, \& {Sterzik}}]{Chini+04}
{Chini}, R., {Hoffmeister}, V., {Kimeswenger}, S., {et~al.} 2004, \nat, 429,
  155

\bibitem[{{Chini} {et~al.}(2006){Chini}, {Hoffmeister}, {Nielbock}, {Scheyda},
  {Steinacker}, {Siebenmorgen}, \& {N{\"u}rnberger}}]{Chini+06}
{Chini}, R., {Hoffmeister}, V.~H., {Nielbock}, M., {et~al.} 2006, \apjl, 645,
  L61

\bibitem[{{Da Rio} {et~al.}(2009){Da Rio}, {Robberto}, {Soderblom}, {Panagia},
  {Hillenbrand}, {Palla}, \& {Stassun}}]{DaRio+09}
{Da Rio}, N., {Robberto}, M., {Soderblom}, D.~R., {et~al.} 2009, \apjs, 183,
  261

\bibitem[{{Da Rio} {et~al.}(2017){Da Rio}, {Tan}, {Covey}, {Cottaar}, {Foster},
  {Cullen}, {Tobin}, {Kim}, {Meyer}, {Nidever}, {Stassun}, {Chojnowski},
  {Flaherty}, {Majewski}, {Skrutskie}, {Zasowski}, \& {Pan}}]{DaRio+17}
{Da Rio}, N., {Tan}, J.~C., {Covey}, K.~R., {et~al.} 2017, \apj, 845, 105

\bibitem[{{Da Rio} {et~al.}(2016){Da Rio}, {Tan}, {Covey}, {Cottaar}, {Foster},
  {Cullen}, {Tobin}, {Kim}, {Meyer}, {Nidever}, {Stassun}, {Chojnowski},
  {Flaherty}, {Majewski}, {Skrutskie}, {Zasowski}, \& {Pan}}]{DaRio+16}
{Da Rio}, N., {Tan}, J.~C., {Covey}, K.~R., {et~al.} 2016, \apj, 818, 59

\bibitem[{{De Pree} {et~al.}(2014){De Pree}, {Peters}, {Mac Low}, {Wilner},
  {Goss}, {Galv{\'a}n-Madrid}, {Keto}, {Klessen}, \& {Monsrud}}]{DePree2014}
{De Pree}, C.~G., {Peters}, T., {Mac Low}, M.-M., {et~al.} 2014, \apjl, 781,
  L36

\bibitem[{{Draine}(2011)}]{Draine2011}
{Draine}, B.~T. 2011, {Physics of the Interstellar and Intergalactic Medium}

\bibitem[{{Duarte-Cabral} {et~al.}(2013){Duarte-Cabral}, {Bontemps}, {Motte},
  {Hennemann}, {Schneider}, \& {Andr{\'e}}}]{Duarte-Cabral+13}
{Duarte-Cabral}, A., {Bontemps}, S., {Motte}, F., {et~al.} 2013, \aap, 558,
  A125

\bibitem[{{Farias} \& {Tan}(2017)}]{FT18}
{Farias}, J.~P. \& {Tan}, J.~C. 2017, ArXiv e-prints

\bibitem[{{Federrath}(2015)}]{Federrath15}
{Federrath}, C. 2015, \mnras, 450, 4035

\bibitem[{{Federrath}(2016)}]{Federrath16}
{Federrath}, C. 2016, \mnras, 457, 375

\bibitem[{{Federrath} {et~al.}(2014){Federrath}, {Schr{\"o}n}, {Banerjee}, \&
  {Klessen}}]{Federrath+14}
{Federrath}, C., {Schr{\"o}n}, M., {Banerjee}, R., \& {Klessen}, R.~S. 2014,
  \apj, 790, 128

\bibitem[{{Felli} {et~al.}(1993){Felli}, {Churchwell}, {Wilson}, \&
  {Taylor}}]{Felli1993}
{Felli}, M., {Churchwell}, E., {Wilson}, T.~L., \& {Taylor}, G.~B. 1993, \aaps,
  98, 137

\bibitem[{{Figer}(2005)}]{Figer05}
{Figer}, D.~F. 2005, \nat, 434, 192

\bibitem[{{For} \& {Bekki}(2017)}]{FB17}
{For}, B.-Q. \& {Bekki}, K. 2017, \mnras, 468, L11

\bibitem[{{Getman} {et~al.}(2014){Getman}, {Feigelson}, \& {Kuhn}}]{Getman+14}
{Getman}, K.~V., {Feigelson}, E.~D., \& {Kuhn}, M.~A. 2014, \apj, 787, 109

\bibitem[{{Gualandris} {et~al.}(2004){Gualandris}, {Portegies Zwart}, \&
  {Eggleton}}]{Gualandris+04}
{Gualandris}, A., {Portegies Zwart}, S., \& {Eggleton}, P.~P. 2004, \mnras,
  350, 615

\bibitem[{{Hacar} {et~al.}(2017{\natexlab{a}}){Hacar}, {Alves}, {Tafalla}, \&
  {Goicoechea}}]{Hacar+17}
{Hacar}, A., {Alves}, J., {Tafalla}, M., \& {Goicoechea}, J.~R.
  2017{\natexlab{a}}, \aap, 602, L2

\bibitem[{{Hacar} {et~al.}(2017{\natexlab{b}}){Hacar}, {Tafalla}, \&
  {Alves}}]{Hacar+17b}
{Hacar}, A., {Tafalla}, M., \& {Alves}, J. 2017{\natexlab{b}}, \aap, 606, A123

\bibitem[{{Hacar} {et~al.}(2013){Hacar}, {Tafalla}, {Kauffmann}, \&
  {Kov{\'a}cs}}]{Hacar+13}
{Hacar}, A., {Tafalla}, M., {Kauffmann}, J., \& {Kov{\'a}cs}, A. 2013, \aap,
  554, A55

\bibitem[{{Hillenbrand} \& {Hartmann}(1998)}]{HH98}
{Hillenbrand}, L.~A. \& {Hartmann}, L.~W. 1998, \apj, 492, 540

\bibitem[{{Hoogerwerf} {et~al.}(2001){Hoogerwerf}, {de Bruijne}, \& {de
  Zeeuw}}]{Hoogerwerf+01}
{Hoogerwerf}, R., {de Bruijne}, J.~H.~J., \& {de Zeeuw}, P.~T. 2001, \aap, 365,
  49

\bibitem[{{Hosokawa} \& {Inutsuka}(2006)}]{Hosokawa2006}
{Hosokawa}, T. \& {Inutsuka}, S.-i. 2006, \apj, 646, 240

\bibitem[{{Hsu} {et~al.}(2012){Hsu}, {Hartmann}, {Allen}, {Hern{\'a}ndez},
  {Megeath}, {Mosby}, {Tobin}, \& {Espaillat}}]{Hsu12}
{Hsu}, W.-H., {Hartmann}, L., {Allen}, L., {et~al.} 2012, \apj, 752, 59

\bibitem[{{Hsu} {et~al.}(2013){Hsu}, {Hartmann}, {Allen}, {Hern{\'a}ndez},
  {Megeath}, {Tobin}, \& {Ingleby}}]{Hsu13}
{Hsu}, W.-H., {Hartmann}, L., {Allen}, L., {et~al.} 2013, \apj, 764, 114

\bibitem[{{Keto} \& {Wood}(2006)}]{Wood2006}
{Keto}, E. \& {Wood}, K. 2006, \apj, 637, 850

\bibitem[{{Kirk} {et~al.}(2016){Kirk}, {Johnstone}, {Di Francesco}, {Lane},
  {Buckle}, {Berry}, {Broekhoven-Fiene}, {Currie}, {Fich}, {Hatchell},
  {Jenness}, {Mottram}, {Nutter}, {Pattle}, {Pineda}, {Quinn}, {Salji}, {Tisi},
  {Hogerheijde}, {Ward-Thompson}, \& {The JCMT Gould Belt Survey
  Team}}]{Kirk16}
{Kirk}, H., {Johnstone}, D., {Di Francesco}, J., {et~al.} 2016, \apj, 821, 98

\bibitem[{{Kirk} \& {Myers}(2011)}]{KM11}
{Kirk}, H. \& {Myers}, P.~C. 2011, \apj, 727, 64

\bibitem[{{Kirk} \& {Myers}(2012)}]{KM12}
{Kirk}, H. \& {Myers}, P.~C. 2012, \apj, 745, 131

\bibitem[{{Klaassen} \& {Wilson}(2007)}]{Klaassen2007}
{Klaassen}, P.~D. \& {Wilson}, C.~D. 2007, \apj, 663, 1092

\bibitem[{{Koen}(2006)}]{Koen06}
{Koen}, C. 2006, \mnras, 365, 590

\bibitem[{{Kounkel} {et~al.}(2017){Kounkel}, {Hartmann}, {Loinard},
  {Ortiz-Le{\'o}n}, {Mioduszewski}, {Rodr{\'{\i}}guez}, {Dzib}, {Torres},
  {Pech}, {Galli}, {Rivera}, {Boden}, {Evans}, {Brice{\~n}o}, \&
  {Tobin}}]{Kounkel+17}
{Kounkel}, M., {Hartmann}, L., {Loinard}, L., {et~al.} 2017, \apj, 834, 142

\bibitem[{{Kraus} {et~al.}(2009){Kraus}, {Weigelt}, {Balega}, {Docobo},
  {Hofmann}, {Preibisch}, {Schertl}, {Tamazian}, {Driebe}, {Ohnaka}, {Petrov},
  {Sch{\"o}ller}, \& {Smith}}]{Kraus+09}
{Kraus}, S., {Weigelt}, G., {Balega}, Y.~Y., {et~al.} 2009, \aap, 497, 195

\bibitem[{{Kroupa}(2000)}]{K2000}
{Kroupa}, P. 2000, \na, 4, 615

\bibitem[{{Kroupa}(2001)}]{Kroupa01}
{Kroupa}, P. 2001, \mnras, 322, 231

\bibitem[{{Kroupa}(2002)}]{Kroupa02b}
{Kroupa}, P. 2002, Science, 295, 82

\bibitem[{{Kroupa}(2008)}]{Kroupa08}
{Kroupa}, P. 2008, in Lecture Notes in Physics, Berlin Springer Verlag, Vol.
  760, The Cambridge N-Body Lectures, ed. S.~J. {Aarseth}, C.~A. {Tout}, \&
  R.~A. {Mardling}, 181

\bibitem[{{Kroupa} {et~al.}(2001){Kroupa}, {Aarseth}, \& {Hurley}}]{KAH}
{Kroupa}, P., {Aarseth}, S., \& {Hurley}, J. 2001, \mnras, 321, 699

\bibitem[{{Kroupa} {et~al.}(1991){Kroupa}, {Gilmore}, \& {Tout}}]{KTG91}
{Kroupa}, P., {Gilmore}, G., \& {Tout}, C.~A. 1991, \mnras, 251, 293

\bibitem[{{Kroupa} {et~al.}(1999){Kroupa}, {Petr}, \&
  {McCaughrean}}]{Kroupa+99}
{Kroupa}, P., {Petr}, M.~G., \& {McCaughrean}, M.~J. 1999, \na, 4, 495

\bibitem[{{Kroupa} {et~al.}(2013){Kroupa}, {Weidner}, {Pflamm-Altenburg},
  {Thies}, {Dabringhausen}, {Marks}, \& {Maschberger}}]{Kroupa13}
{Kroupa}, P., {Weidner}, C., {Pflamm-Altenburg}, J., {et~al.} 2013, {The
  stellar and sub-stellar initial mass function of simple and composite
  populations}, ed. T.~D. {Oswalt} \& G.~{Gilmore}, 115

\bibitem[{{Kuruwita} {et~al.}(2017){Kuruwita}, {Federrath}, \&
  {Ireland}}]{Kuruwita+17}
{Kuruwita}, R.~L., {Federrath}, C., \& {Ireland}, M. 2017, \mnras, 470, 1626

\bibitem[{{Lada}(2010)}]{Lada10}
{Lada}, C.~J. 2010, Philosophical Transactions of the Royal Society of London
  Series A, 368, 713

\bibitem[{{Lada} \& {Lada}(2003)}]{LL03}
{Lada}, C.~J. \& {Lada}, E.~A. 2003, \araa, 41, 57

\bibitem[{{Lane} {et~al.}(2016){Lane}, {Kirk}, {Johnstone}, {Mairs}, {Di
  Francesco}, {Sadavoy}, {Hatchell}, {Berry}, {Jenness}, {Hogerheijde},
  {Ward-Thompson}, \& {The JCMT Gould Belt Survey Team}}]{Lane+16}
{Lane}, J., {Kirk}, H., {Johnstone}, D., {et~al.} 2016, \apj, 833, 44

\bibitem[{{Luhman} {et~al.}(2017){Luhman}, {Robberto}, {Tan}, {Andersen},
  {Giulia Ubeira Gabellini}, {Manara}, {Platais}, \& {Ubeda}}]{Luhman+17}
{Luhman}, K.~L., {Robberto}, M., {Tan}, J.~C., {et~al.} 2017, \apjl, 838, L3

\bibitem[{{Machida} \& {Matsumoto}(2012)}]{MM12}
{Machida}, M.~N. \& {Matsumoto}, T. 2012, \mnras, 421, 588

\bibitem[{{Marks} \& {Kroupa}(2012)}]{MK12}
{Marks}, M. \& {Kroupa}, P. 2012, \aap, 543, A8

\bibitem[{{Maschberger} \& {Clarke}(2011)}]{MC11}
{Maschberger}, T. \& {Clarke}, C.~J. 2011, \mnras, 416, 541

\bibitem[{{Mc Leod} {et~al.}(2016){Mc Leod}, {Weilbacher}, {Ginsburg}, {Dale},
  {Ramsay}, \& {Testi}}]{McLeod+16}
{Mc Leod}, A.~F., {Weilbacher}, P.~M., {Ginsburg}, A., {et~al.} 2016, \mnras,
  455, 4057

\bibitem[{{Megeath} {et~al.}(2016){Megeath}, {Gutermuth}, {Muzerolle},
  {Kryukova}, {Hora}, {Allen}, {Flaherty}, {Hartmann}, {Myers}, {Pipher},
  {Stauffer}, {Young}, \& {Fazio}}]{Megeath16}
{Megeath}, S.~T., {Gutermuth}, R., {Muzerolle}, J., {et~al.} 2016, \aj, 151, 5

\bibitem[{{Menten} {et~al.}(2007){Menten}, {Reid}, {Forbrich}, \&
  {Brunthaler}}]{Menten+07}
{Menten}, K.~M., {Reid}, M.~J., {Forbrich}, J., \& {Brunthaler}, A. 2007, \aap,
  474, 515

\bibitem[{{Moe} \& {Di Stefano}(2017)}]{MDiS17}
{Moe}, M. \& {Di Stefano}, R. 2017, \apjs, 230, 15

\bibitem[{{Nielbock} {et~al.}(2007){Nielbock}, {Chini}, {Hoffmeister},
  {Scheyda}, {Steinacker}, {N{\"u}rnberger}, \& {Siebenmorgen}}]{Nielbock+07}
{Nielbock}, M., {Chini}, R., {Hoffmeister}, V.~H., {et~al.} 2007, \apjl, 656,
  L81

\bibitem[{{N{\"u}rnberger} {et~al.}(2007){N{\"u}rnberger}, {Chini},
  {Eisenhauer}, {Kissler-Patig}, {Modigliani}, {Siebenmorgen}, {Sterzik}, \&
  {Szeifert}}]{Nuernberger+07}
{N{\"u}rnberger}, D.~E.~A., {Chini}, R., {Eisenhauer}, F., {et~al.} 2007, \aap,
  465, 931

\bibitem[{{O'Dell}(1998)}]{ODell98}
{O'Dell}, C.~R. 1998, \aj, 115, 263

\bibitem[{{O'dell}(2001)}]{Odell2001}
{O'dell}, C.~R. 2001, \araa, 39, 99

\bibitem[{{O'Dell} {et~al.}(2009){O'Dell}, {Henney}, {Abel}, {Ferland}, \&
  {Arthur}}]{ODell+09}
{O'Dell}, C.~R., {Henney}, W.~J., {Abel}, N.~P., {Ferland}, G.~J., \& {Arthur},
  S.~J. 2009, \aj, 137, 367

\bibitem[{{Oey} \& {Clarke}(2005)}]{OC05}
{Oey}, M.~S. \& {Clarke}, C.~J. 2005, \apjl, 620, L43

\bibitem[{{Oh} \& {Kroupa}(2016)}]{OK16}
{Oh}, S. \& {Kroupa}, P. 2016, \aap, 590, A107

\bibitem[{{Oh} {et~al.}(2015){Oh}, {Kroupa}, \& {Pflamm-Altenburg}}]{OKP15}
{Oh}, S., {Kroupa}, P., \& {Pflamm-Altenburg}, J. 2015, \apj, 805, 92

\bibitem[{{Peters} {et~al.}(2010{\natexlab{a}}){Peters}, {Banerjee}, {Klessen},
  {Mac Low}, {Galv{\'a}n-Madrid}, \& {Keto}}]{Peters2010a}
{Peters}, T., {Banerjee}, R., {Klessen}, R.~S., {et~al.} 2010{\natexlab{a}},
  \apj, 711, 1017

\bibitem[{{Peters} {et~al.}(2010{\natexlab{b}}){Peters}, {Mac Low}, {Banerjee},
  {Klessen}, \& {Dullemond}}]{Peters2010b}
{Peters}, T., {Mac Low}, M.-M., {Banerjee}, R., {Klessen}, R.~S., \&
  {Dullemond}, C.~P. 2010{\natexlab{b}}, \apj, 719, 831

\bibitem[{{Pflamm-Altenburg} \& {Kroupa}(2006)}]{PAK06}
{Pflamm-Altenburg}, J. \& {Kroupa}, P. 2006, \mnras, 373, 295

\bibitem[{{Pflamm-Altenburg} \& {Kroupa}(2007)}]{PAK07}
{Pflamm-Altenburg}, J. \& {Kroupa}, P. 2007, \mnras, 375, 855

\bibitem[{{Pflamm-Altenburg} \& {Kroupa}(2009)}]{PAK09}
{Pflamm-Altenburg}, J. \& {Kroupa}, P. 2009, \mnras, 397, 488

\bibitem[{{Pflamm-Altenburg} \& {Kroupa}(2010)}]{Pflamm10}
{Pflamm-Altenburg}, J. \& {Kroupa}, P. 2010, \mnras, 404, 1564

\bibitem[{{Portegies Zwart} {et~al.}(2009){Portegies Zwart}, {McMillan},
  {Harfst}, {Groen}, {Fujii}, {Nuall{\'a}in}, {Glebbeek}, {Heggie}, {Lombardi},
  {Hut}, {Angelou}, {Banerjee}, {Belkus}, {Fragos}, {Fregeau}, {Gaburov},
  {Izzard}, {Juri{\'c}}, {Justham}, {Sottoriva}, {Teuben}, {van Bever},
  {Yaron}, \& {Zemp}}]{Portegies09}
{Portegies Zwart}, S., {McMillan}, S., {Harfst}, S., {et~al.} 2009, \na, 14,
  369

\bibitem[{{Poveda} {et~al.}(2005){Poveda}, {Allen}, \&
  {Hern{\'a}ndez-Alc{\'a}ntara}}]{Poveda+05}
{Poveda}, A., {Allen}, C., \& {Hern{\'a}ndez-Alc{\'a}ntara}, A. 2005, \apjl,
  627, L61

\bibitem[{{Prosser} {et~al.}(1994){Prosser}, {Stauffer}, {Hartmann},
  {Soderblom}, {Jones}, {Werner}, \& {McCaughrean}}]{Prosser+94}
{Prosser}, C.~F., {Stauffer}, J.~R., {Hartmann}, L., {et~al.} 1994, \apj, 421,
  517

\bibitem[{{Ram{\'{\i}}rez Alegr{\'{\i}}a} {et~al.}(2016){Ram{\'{\i}}rez
  Alegr{\'{\i}}a}, {Borissova}, {Chen{\'e}}, {Bonatto}, {Kurtev}, {Amigo},
  {Kuhn}, {Gromadzki}, \& {Carballo-Bello}}]{Ramirez16}
{Ram{\'{\i}}rez Alegr{\'{\i}}a}, S., {Borissova}, J., {Chen{\'e}}, A.-N.,
  {et~al.} 2016, \aap, 588, A40

\bibitem[{{Sana} {et~al.}(2012){Sana}, {de Mink}, {de Koter}, {Langer},
  {Evans}, {Gieles}, {Gosset}, {Izzard}, {Le Bouquin}, \&
  {Schneider}}]{Sana+12}
{Sana}, H., {de Mink}, S.~E., {de Koter}, A., {et~al.} 2012, Science, 337, 444

\bibitem[{{Scally} \& {Clarke}(2002)}]{Scally+02}
{Scally}, A. \& {Clarke}, C. 2002, \mnras, 334, 156

\bibitem[{{Scally} {et~al.}(2005){Scally}, {Clarke}, \&
  {McCaughrean}}]{Scally+05}
{Scally}, A., {Clarke}, C., \& {McCaughrean}, M.~J. 2005, \mnras, 358, 742

\bibitem[{{Schneider} {et~al.}(2012){Schneider}, {Csengeri}, {Hennemann},
  {Motte}, {Didelon}, {Federrath}, {Bontemps}, {Di Francesco}, {Arzoumanian},
  {Minier}, {Andr{\'e}}, {Hill}, {Zavagno}, {Nguyen-Luong}, {Attard},
  {Bernard}, {Elia}, {Fallscheer}, {Griffin}, {Kirk}, {Klessen}, {K{\"o}nyves},
  {Martin}, {Men'shchikov}, {Palmeirim}, {Peretto}, {Pestalozzi}, {Russeil},
  {Sadavoy}, {Sousbie}, {Testi}, {Tremblin}, {Ward-Thompson}, \&
  {White}}]{Schneider+12}
{Schneider}, N., {Csengeri}, T., {Hennemann}, M., {et~al.} 2012, \aap, 540, L11

\bibitem[{{Smith} {et~al.}(2016){Smith}, {Glover}, {Klessen}, \&
  {Fuller}}]{Smith+16}
{Smith}, R.~J., {Glover}, S.~C.~O., {Klessen}, R.~S., \& {Fuller}, G.~A. 2016,
  \mnras, 455, 3640

\bibitem[{{Spitzer}(1978)}]{Spitzer1978}
{Spitzer}, L. 1978, {Physical processes in the interstellar medium}

\bibitem[{{Stephens} {et~al.}(2017){Stephens}, {Gouliermis}, {Looney},
  {Gruendl}, {Chu}, {Weisz}, {Seale}, {Chen}, {Wong}, {Hughes}, {Pineda},
  {Ott}, \& {Muller}}]{Stephens+17}
{Stephens}, I.~W., {Gouliermis}, D., {Looney}, L.~W., {et~al.} 2017, \apj, 834,
  94

\bibitem[{Sternberg {et~al.}(2003)Sternberg, Hoffmann, \&
  Pauldrach}]{Sternberg2003}
Sternberg, A., Hoffmann, T.~L., \& Pauldrach, A. W.~A. 2003, The Astrophysical
  Journal, 599, 1333

\bibitem[{{Stutz} \& {Gould}(2016)}]{SG16}
{Stutz}, A.~M. \& {Gould}, A. 2016, \aap, 590, A2

\bibitem[{{Tan}(2004)}]{Tan04}
{Tan}, J.~C. 2004, \apjl, 607, L47

\bibitem[{{Tenorio-Tagle}(1979)}]{Tenorio-Tagle1979}
{Tenorio-Tagle}, G. 1979, \aap, 71, 59

\bibitem[{{{\v S}ubr} {et~al.}(2012){{\v S}ubr}, {Kroupa}, \&
  {Baumgardt}}]{Subr+12}
{{\v S}ubr}, L., {Kroupa}, P., \& {Baumgardt}, H. 2012, \apj, 757, 37

\bibitem[{{Weidner} \& {Kroupa}(2004)}]{WK04}
{Weidner}, C. \& {Kroupa}, P. 2004, \mnras, 348, 187

\bibitem[{{Weidner} \& {Kroupa}(2006)}]{Weidner06}
{Weidner}, C. \& {Kroupa}, P. 2006, \mnras, 365, 1333

\bibitem[{{Weidner} {et~al.}(2013){Weidner}, {Kroupa}, \&
  {Pflamm-Altenburg}}]{Weidner13}
{Weidner}, C., {Kroupa}, P., \& {Pflamm-Altenburg}, J. 2013, \mnras, 434, 84

\bibitem[{{Wen} \& {O'dell}(1995)}]{Wen1995}
{Wen}, Z. \& {O'dell}, C.~R. 1995, \apj, 438, 784

\bibitem[{{Whitworth}(1979)}]{Whitworth1979}
{Whitworth}, A. 1979, \mnras, 186, 59

\bibitem[{{Wood} \& {Churchwell}(1989)}]{Wood1989}
{Wood}, D.~O.~S. \& {Churchwell}, E. 1989, \apjs, 69, 831

\bibitem[{{Wuchterl} \& {Tscharnuter}(2003)}]{WT03}
{Wuchterl}, G. \& {Tscharnuter}, W.~M. 2003, \aap, 398, 1081

\bibitem[{{Yan} {et~al.}(2017){Yan}, {Jerabkova}, \& {Kroupa}}]{Yan+17}
{Yan}, Z., {Jerabkova}, T., \& {Kroupa}, P. 2017, \aap, 607, A126

\end{thebibliography}
\end{document}